\documentclass{ws-procs9x6}            
\begin{document}
\title{Weyl Cosmology}

\author{V. A. Berezin$^*$ and V. I. Dokuchaev$^{**}$}

\address{Institute for Nuclear Research of the Russian Academy of Sciences,\\
 Moscow, 117312, prospekt 60-letiya Oktyabrya 7a, Russia City \\
$^*$E-mail: berezin@inr.ac.ru, $^{**}$E-mail: dokuchaev@inr.ac.ru}

\begin{abstract}
The homogeneous and isotropic cosmological model in the Weyl conformal geometry is considered. We showed that, despite the conformal invariance, the dust matter is allowed in such a universe. It is shown that the number of dust particles is not conserved, i.\,e., they are continuously produced. The general form of the law for their creation is found.
\end{abstract}

\keywords{Gravitation, Weyl geometry, conformal symmetry, cosmology}

\bodymatter

\section{Introduction}

In the present paper we assume that the universe, as a whole, is conformal invariant. It is quite reasonable if the classical universe was created by tunneling process from some quantum state, because the more symmetry --- the easier birth. The same point of view was advocated by Roger Penrose\cite{Penrose,Penrose2,Penrose3} and Gerard 't Hooft\cite{Hooft,Hooft2,Hooft3}. It is for this very reason that we consider the conformal gravity and the homogeneous and isotropic cosmological model. In the framework of Riemannian geometry any homogeneous and isotropic metric is a vacuum solution to the conformal gravity equations\cite{bde16,bdes21}, i.\,e., the universe cannot be filled with the matter field. That why we turned to the Weyl geometry\cite{Weyl,Weyl2}. While the Riemanian geometry is purely metric one, the Weyl geometry contains in addition, the vector field. Originally, Hermann Weyl identified  the latter with the electromagnetic potential. We will not do this and consider the Weyl vector just as the part of the geometry needed to provide the conformal invariance to the gravitational theory.

\section{Mathematical preliminaries of Weyl geometry}

Everybody knows that any specific differential geometry is completely defined by some metric tensor $g_{\mu\nu}$ and connections $\Gamma^\lambda_{\mu\nu}$. The metric tensor gives us the interval $ds$ between the neighboring points
\begin{equation}
	ds^2=g_{\mu\nu}(x)dx^\mu dx^\nu,
	\label{ds}
\end{equation}
while the connections are needed for construction of covariant derivatives $\nabla_\mu$. By definition, for scalars
\begin{equation}
	\nabla_\mu\varphi(x)=\varphi_{,\mu}(x)
	\label{varphi}
\end{equation}
(coma ``,''denotes a partial derivative), for vectors $l^\lambda(l_\lambda)$
\begin{equation}
	\nabla_\mu l^\lambda =l^\lambda_{,\mu}+\Gamma^\lambda_{\sigma\mu}l^\sigma
	\label{vector1}
\end{equation}
\begin{equation}
	\nabla_\mu l_\lambda =l^\lambda_{,\mu}-\Gamma^\sigma_{\lambda\mu}l_\sigma
	\label{vector2}
\end{equation}
(the change in signs is dictated by the Leibniz' rule, since $l^\mu l_\mu$ is a scalar), and for tensors (say $A^{\mu\nu}$)  one has (again because of the Leibniz' rule)
\begin{equation}
	\nabla_\mu A^{\nu\lambda} =A^{\nu\lambda}_{\phantom{\mu},\mu} +\Gamma^\lambda_{\sigma\mu}A^{\nu\sigma}
	\label{tensors}
\end{equation}
(note, that the position of the index $\mu$ in the above formulas is, in general, important).

Given the connections, one can construct the Riemann tensor
\begin{equation}
	R^{\mu}_{\phantom{\mu}\nu\lambda\sigma}=\frac{\partial \Gamma^\mu_{\nu\sigma}}{\partial x^\lambda}-\frac{\partial \Gamma^\mu_{\nu\lambda}}{\partial x^\sigma}+\Gamma^\mu_{\varkappa\lambda}\Gamma^\varkappa_{\nu\sigma}-\Gamma^\mu_{\varkappa\sigma}\Gamma^\varkappa_{\nu\lambda},
	\label{curvature}
\end{equation}
which reflects the properties of neighborhoods of the point $x$, its convolutions, Ricci tensor $R_{\mu\nu}$ and curvature tensor $R$,
\begin{equation}
	R_{\mu\nu}=R^{\lambda}_{\mu\lambda\nu},
	\label{ricci}
\end{equation}
\begin{equation}
	R=g^{\mu\lambda}R_{\mu\lambda}.
	\label{curvature}
\end{equation}
It appears that connections $\Gamma^\lambda_{\mu\nu}$ can be calculated provided three tensors are given: the metric tensor $g_{\mu\nu}$ the torsion tensor  $S^\lambda_{\mu\nu}$,
\begin{equation}
	S^\lambda_{\mu\nu}=\Gamma^\lambda_{\mu\nu} -\Gamma^\lambda_{\nu\mu},
	\label{torsion}
\end{equation}
and the so called nonmetricity $Q_{\mu\nu\lambda}$, 
\begin{equation}
	Q_{\mu\nu\lambda}=\nabla_\mu  g_{\nu\lambda}.
	\label{Q}
\end{equation}
Namely, 
\begin{equation}
	\Gamma^\lambda_{\mu\nu}=C^\lambda_{\mu\nu}+K^\lambda_{\mu\nu}
	+L^\lambda_{\mu\nu},
	\label{Gamma3}
\end{equation}
\begin{equation}
	C^\lambda_{\mu\nu}=\frac{1}{2}g^{\lambda\varkappa}(g_{\varkappa\mu,\nu}+g_{\varkappa\nu,\mu}-g_{\mu\nu,\varkappa}),
	\label{c}
\end{equation}
\begin{equation}
	K^\lambda_{\phantom{1}\mu\nu}=\frac{1}{2}(S^\lambda_{\phantom{1}\mu\nu}-S^{\phantom{1}\lambda}_{\mu\phantom{1}\nu}-S^{\phantom{1}\lambda}_{\nu\phantom{1}\mu}),
	\label{K}
\end{equation}
\begin{equation}
	L^\lambda_{\mu\nu}=\frac{1}{2}(Q^\lambda_{\phantom{1}\mu\nu} -Q^{\phantom{1}\lambda}_{\mu\phantom{1}\nu}-Q^{\phantom{1}\lambda}_{\nu\phantom{1}\mu}).
	\label{L}
\end{equation}
Clearly, $C^\lambda_{\mu\nu}$ are familiar Christoffel symbols.

The Riemannian geometry is the simplest case: both the torsion tensor and the nonmetricity are zero:
\begin{equation}
	S^\lambda_{\mu\nu}=0, \quad Q_{\mu\nu\lambda}=0.
	\label{Riemannian}
\end{equation}
Thus,
\begin{equation}
	\Gamma^\lambda_{\mu\nu}=\Gamma^\lambda_{\nu\mu}=C^\lambda_{\mu\nu}.
	\label{GammaR}
\end{equation}
and the Riemannian geometry geometry is purely metrical.

The Weyl geometry differs from the Riemannian one that, in addition to the metric tensor $g_{\mu\nu}$, it contains the so called Weyl vector  (1-form) $A_\mu$, introduced by the relation
\begin{equation}
	\nabla_\mu g_{\nu\lambda}=A_\mu g_{\nu\lambda},
	\label{A}
\end{equation}
that is, the nonmetricity tensor is no more zero, but equals $Q_{\mu\nu\lambda}=A_\mu g_{\nu\lambda}$, the torsion tensor $S^\lambda_{\mu\nu}$ is still being zero.

Hence, in Weyl geometry
\begin{equation}
	\Gamma^\lambda_{\mu\nu}=C^\lambda_{\mu\nu}+W^\lambda_{\mu\nu},
	\label{ WeylGamma}
\end{equation}
\begin{equation}
	W^\lambda_{\mu\nu}=-\frac{1}{2}(A_\mu \delta^\lambda_\nu + A_\nu \delta^\lambda_\mu - A^\lambda g_{\mu\nu}).
	\label{W}
\end{equation}

\section{Weyl gravity and Weyl cosmology. Field equations.}

One hundred years ago Hermann Weyl made an attempt to construct the unified geometric theory of gravitational and electromagnetic interactions. He noticed that Maxwell equations are invariant under both gauge and local conformal transformations provided the electromagnetic stress tensor $F_{\mu\nu}= A_{\nu,\mu}-A_{\mu,\nu}$, built of the vector potential $A_\mu$, remains untouched. But, the gauge transformation
\begin{equation}
	A_\mu \quad \rightarrow \quad A_\mu+\alpha(x)_{,\mu},
	\label{gaugeA}
\end{equation}
($\alpha(x)$ is some scalar field) is associated with the electromagnetic theory, while the conformal transformation
\begin{equation}
	g_{\mu\nu}=\Omega^2\hat g_{\mu\nu}
	\label{conformalg}
\end{equation}
is purely geometrical. H.\,Weyl decided to link them to each other, this was the first step to the unification. He also claimed that the whole theory also should be conformal invariant. Weyl considered the parallel transport of vectors. It is well known that in the Riemannian geometry the length of vectors remains constant during such a process. But, in the unified theory this is not true since the electromagnetic force interacts with the charged particles in the rods (brilliant physical idea for the mathematician) He managed to find  the corresponding connections and discovered that they are conformal invariant if, under the conformal transformation with the conformal factor $\Omega(x)$, the corresponding gauge transformation of vector potential $A_\mu$ is
\begin{equation}
	A_\mu=\hat A_\mu+2\frac{\Omega_{,\mu}}{\Omega} \quad  
	\label{conformalA}
\end{equation}
(i.\,e.,\, $\alpha(x)=\log[\Omega^2]$).

With this choice of gauge transformation of the Weyl vector $A_\mu$ one has
\begin{equation}
	R^\mu_{\phantom{1}\nu\lambda\sigma}=\hat R^\mu_{\phantom{1}\nu\lambda\sigma}, 
	\label{conformalRi2}
\end{equation}
\begin{equation}
	R_{\mu\nu}=\hat R_{\mu\nu},
	\label{conformalRicci}
\end{equation}
\begin{equation}
	F_{\mu\nu}=\hat F_{\mu\nu}.
	\label{conformalF}
\end{equation}

Dealing with the quadratic (at most) combinations, one can construct the only possible invariant action integral for the Weyl gravity,
\begin{equation}
	S_{\rm  W}=\int\!{\cal L_{\rm W}}\sqrt{-g}\,d^4x , \quad \frac{\delta S_{\rm  W}}{\delta\Omega}=0,
	\label{WeyalAction}
\end{equation}
\begin{equation}
{\cal L_{\rm  W}} =\alpha_1  R_{\mu\nu\lambda\sigma}R^{\mu\nu\lambda\sigma}
+\alpha_2R_{\mu\nu}R^{\mu\nu}+\alpha_3R^2+ \alpha_4 F_{\mu\nu}F^{\mu\nu}
	\label{WLagr}
\end{equation}
with constant $(\alpha_1,\alpha_2,\alpha_3,\alpha_4$). The dynamical variables in the Weyl gravity are $g_{\mu\nu}(x)$ and $A_\mu(x)$. 

We do not want to identify the Weyl vector $A_\mu(x)$ with the electromagnetic potential. Instead, we will consider it just as a part of Weyl geometry.

The total action integral is
\begin{equation}
S_{\rm tot}=S_{\rm W}+S_{\rm m}, \quad .
	\label{total}
\end{equation}
\begin{equation}
S_{\rm  m}=\int\!{\cal L_{\rm m}}\sqrt{-g}\,d^4x,
	\label{Sm}
\end{equation}
where $L_{\rm m}$ is a Lagrangian of the matter fields. Very important to realize that $S_{\rm  m}$ is not necessary conformal invariant. But, its variation $\delta S_{\rm m}$, must obey such a requirement. By definition,
\begin{eqnarray}
\delta S_{\rm m}\stackrel{\mathrm{def}}{=} &-&\frac{1}{2}\int\! T^{\mu\nu}(\delta g_{\mu\nu})\sqrt{-g}\,d^4x
-\!\int\!\! G^\mu(\delta A_\mu)\sqrt{-g}\,d^4x 
\nonumber\\
&+&\int\!\frac{\delta \cal L_{\rm  m}}{\delta\Psi} (\delta \Psi)\sqrt{-g}\,d^4x=0.
	\label{Sm}
\end{eqnarray}
Here $T^{\mu\nu}$ is the energy momentum tensor of the matter fields, $G^\mu$ can be called ``{\it the Weyl current}'', and $\Psi$ is the collective dynamical variable describing the dynamics of matter fields. As usual ${\delta \cal L_{\rm  m}}/\delta\Psi =0$. Under the infinitesimal change, $\delta\Omega$, of the conformal factor,
\begin{equation}
\delta g_{\mu\nu}=\frac{2}{\Omega}g_{\mu\nu}(\delta\Omega),
\label{deltag}
\end{equation}
\begin{equation}
\delta A_\mu=2(\delta(\log\Omega)_{,\mu},
\label{deltaA}
\end{equation}
hence
\begin{equation}
\delta S_{\rm m}= -\!\int\! T^{\mu\nu}g_{\mu\nu}\left(\frac{\delta\Omega}{\Omega}\right)\sqrt{-g}\,d^4x
-\!\int\!\! 2G^\mu(\delta(\log\Omega))_{,\mu}\sqrt{-g}\,d^4x.
	\label{deltaA}
\end{equation}
Getting rid of the partial derivatives in the second integral one has the following self-consistency condition
\begin{equation}
2G^\mu_{;\mu}=Trace[T^{\mu\nu}].
	\label{Trace}
\end{equation}
Here  the semicolon ``;'' denotes the metric covariant derivative, that is with Christoffel symbols. This condition is supplement to the field equations.

Let us now come to cosmology. For simplicity, we consider here only homogeneous and isotropic space-times with the Robertson-Walker line element
\begin{equation}
ds^2=dt^2-a^2(t)\left(\frac{dr^2}{1-kr^2}+r^2(d\theta^2+\sin^2\theta d\varphi^2) \right), \quad  k=0,\pm1.
	\label{RW}
\end{equation}
Such a high symmetry dictates 
\begin{equation}
A_mu=(A(t),0,0,0), 	\label{At}
\end{equation}
hence
\begin{equation}
	F_{\mu\nu}\equiv0. 	\label{F0}
\end{equation}
Also
\begin{equation}
T^\mu_\nu=(T^0_0(t),T^1_1(t)=T^2_2=T^3_3). 	\label{T0}
\end{equation}
By suitable choice of the conformal factor $\Omega(t)$ one can always make
\begin{equation}
A_0(t)=0, 	\label{A0t}
\end{equation}
we will call it ``{\it the special gauge\/}''.

Warning: it is not allowed to put $A_\mu=0$ before the variation  procedure, since $\delta A_\mu\neq0$. Keeping this in mind we obtained the following set of field equations in our special gauge,
\begin{equation}
\delta A_\mu:  \quad 	-6\gamma\dot R=G^0, \quad 	 \gamma=\frac{1}{3}(\alpha_1+\alpha_2+3\alpha_3), 	
\label{eq1}
\end{equation}
\begin{equation}
R=-6\left(\frac{\ddot a}{a}+\frac{\dot a^2+k}{a^2}\right), 	
	\label{eq2}
\end{equation}
\begin{equation}
\delta g_{\mu\nu}:  \quad 	-12\gamma{\Bigl\{ }\frac{\dot a}{a}\dot R+ R\left(\frac{R}{12}+\frac{\dot a^2+k}{a^2}\right){\!\Bigl\}}=T^0_0, 	
	\label{eq3}
\end{equation}
\begin{equation}
-4\gamma{\Bigl\{ }\ddot R+2\frac{\dot a}{a}\dot R
- R\left(\frac{R}{12}+\frac{\dot a^2+k}{a^2}\right){\!\Bigl\}}=T^1_1.
	\label{eq4}
\end{equation}
One should add to these equations the self-consistency condition which looks now as follows
\begin{equation}
2\frac{(G^0a^3)^{\dot{}}}{a^3}=T^0_0+3T^1_1.
	\label{selfcons}
\end{equation}
In our special gauge the energy-momentum tensor $T^\mu_\nu$ is automatically conservative,
\begin{equation}
\frac{(a^3T^0_0)^{\dot{}}}{a^3}=3\frac{\dot a}{a}T^1_1.
	\label{selfcons2}
\end{equation}
It easy to see that the self-consistency condition in the above form is just the consequence of other field equations.

Let us consider first the case $G^0=0$, that is , the situation when the matter fields do not interact directly with the Weyl vector $A_mu$. Then, 
$\dot R=0 \quad \Rightarrow \quad R=R_0=const$, and the general solution is
\begin{equation}
\dot a^2+k=-\frac{R_0}{12}a^2+\frac{Q_0}{a^2},
	\label{gensol2}
\end{equation}
where $Q_0$ is an integration constant. Then for the empty universe one has $R_0 Q_0=0$, and we arrive at the following vacuum solutions,
\begin{eqnarray}
	\hspace{-1.6cm}Q_0=0: \quad R_0<0\quad &\Rightarrow& \quad \mbox{de Sitter}, \\
	R_0>0\quad &\Rightarrow& \quad \mbox{AdS}, \\
	R_0=0\quad &\Rightarrow& \quad \mbox{Milne universes}.
	\label{gensol3}
\end{eqnarray}
\begin{eqnarray}
R_0=0: \quad &&\!\!4Q_0(t-t_0)^2=a^2, \quad  k=0   \quad (Q_0>0), \\
	  &&\!\!a^2+(t-t_0)^2=Q_0, \quad k=\pm1   \quad (Q_0>0), \\
	 &&\!\!(t-t_0)^2-a^2, \quad k=-1.
	\label{gensol3}
\end{eqnarray}
There exists also the non-vacuum solution with constant curvature scalar. It represents the universe filled with radiation. The final expression includes elliptic integrals and looks not very pleasant. We do not want to reproduce it here.

\section{Perfect Fluid Dynamics. Particle number non-conservation. Dust matter.}

Let us turn now to the most interesting case $G^0\neq0$. This means that the direct interaction of the manner fields with the Weyl vector $A^\mu$ should be somehow incorporated into the matter Lagrangian. 

As a valuable example we consider here the dust matter. In order to have an idea how the needed interaction could be arranged. let us consider first the single particle moving in some given gravitational field (i.\,e., $g_{\mu\nu}(x)$ and $A_\mu(x)$ are fixed).

It is well known that in the Riemannian geometry the only possible action integral for the single particle is
\begin{equation}
S_{\rm part} =  -m\!\int\!ds=  -m\!\int\!\sqrt{u^\mu u_\mu}d\tau,
	\label{only}
\end{equation}
where $m$ and $\tau$ are, its mass and four velocity, correspondingly. In the Weyl geometry, however, there is yet another invariant, $B$,
\begin{equation}
B=A^\mu u_\mu, 	\label{B}
\end{equation}
and the general form for the action integral is
\begin{equation}
S_{\rm part} =\!\int\!\!\Phi_1(B)\sqrt{u^\mu u_\mu}d\tau \! +\!\int\!\!\Phi_2(B)d\tau.
	\label{part}
\end{equation}
The dynamical variable in this action integral is still the particle trajectory 
$x^\mu(\tau)$ ($u^\mu=dx^\mu/d\tau$). The modified Lagrange equation becomes now
\begin{equation}
-\Phi_1u_{\lambda;\mu}u^\mu -\Phi_1'B_{,\mu}u_\lambda u^\mu - (\Phi_1''+\Phi_2'')B_{,\mu}u_\lambda u^\mu
= (\Phi_1'+\Phi_2')F_{\lambda\mu}u^\mu, 
	\label{Lagrange2}
\end{equation}
where prime ``,'' denotes the derivatives in $B$. Contraction of the above equation with $u^\mu$ gives us the consistency relation 
\begin{equation}
	(B_{,\mu}u^\mu)(\Phi_1'+B(\Phi_1''+\Phi_2''))=0.
	\label{relation}
\end{equation}
If $B_{,\mu}u^\mu=0$, the invariant $B$ is constant along the particle trajectory, and we arrive at the modified Lorentz force.

In the Riemannian geometry the action integral for the perfect fluid can be written in the form
\begin{eqnarray}
	S_{\rm m} &=&  -\!\!\int\!\varepsilon\sqrt{-g}\,d^4x + \!\!\int\!\lambda_0(u_\mu u^\mu-1)\sqrt{-g}\,d^4x \nonumber \\ 
	&&+\!\!\int\!\lambda_1(n u^\mu)_{;\mu}\sqrt{-g}\,d^4x + \!\!\int\!\lambda_2 X_{,\mu}u^\mu\sqrt{-g}\,d^4x.
	\label{SmPerfect} 
\end{eqnarray}
The dynamical variables are particle density $n(x)$, four-velocity $u^\mu(x)$, and auxiliary variable $X(x)$, $\varepsilon(X,x)$ is the invariant energy density, and $\lambda_0$, $\lambda_1$, and $\lambda_2$ are Lagrange multipliers for the ``{\it correct}'' four-velocity normalization, particle number conservation and numbering the particle trajectories, respectively\cite{Ray,Berezin,Ber14}. Naturally, modifications can be made only in the first and third terms. For the dust matter it is obvious that
\begin{equation}
n \quad \rightarrow \quad \varphi(B)n.
	\label{n}
\end{equation}
Let us now suppose that the particle density is conserved. Then, as can be easily shown,
\begin{equation}
T^\mu_\nu=\alpha\varphi(B)nu^\mu u_\nu, \quad \alpha=const,
	\label{T}
\end{equation}
\begin{equation}
Trace[T^{\mu\nu}]=\alpha\varphi(B)n.
	\label{Trace}
\end{equation}
In our special gauge ($B=0$)
\begin{equation}
G^0=\alpha\varphi'(0)n
	\label{G0}
\end{equation}
and self consistency condition gives us
\begin{equation}
\alpha\varphi'(0)\frac{(a^3n)^{\dot{}}}{a^3}=0=\alpha\varphi(0)n, 
	\label{self}
\end{equation}
i.\,e., $n=0$. Thus we arrive at the very important conclusion, that in Weyl geometry the dust particle creation is inevitable even in the homogeneous and isotropic universe. What could be the rate of creation?

The rate of particle production should be some function of invariants constructed from the geometric quantities,
\begin{equation}
	(nu^\mu)_{;\mu}=\Phi(inv).
	\label{inv}
\end{equation}
Let us have a look at the behavior of this under the conformal transformation,
\begin{equation}
	n=\frac{\hat n}{\Omega^4}, \quad u^\mu=\frac{1}{\Omega}\hat u^\mu, \quad \sqrt{-g}=\Omega^4\sqrt{-\hat g},
	\label{intxi4b} 
\end{equation}
we find that the left-hand side is conformal invariant. With the quadratic terms (at most) in the curvatures one may have only the following linear combination
\begin{equation}
\Phi=\alpha'_1  R_{\mu\nu\lambda\sigma}R^{\mu\nu\lambda\sigma}
	+\alpha'_2R_{\mu\nu}R^{\mu\nu}+ \alpha'_3R^2+ \alpha'_4 F_{\mu\nu}F^{\mu\nu}.
	\label{WLagr}
\end{equation}
In the limit $A_\mu=0$ it is reduced to the 
\begin{equation}
\Phi=\alpha  C_{\mu\nu\lambda\sigma}C^{\mu\nu\lambda\sigma}=\alpha C^2,
	\label{WLagr}
\end{equation}
where $C_{\mu\nu\lambda\sigma}$ is the famous Weyl tensor. This result was obtained long ago by Ya. B. Zel'dovich and A. A. Starobinskii\cite{ZeldStar71} for the special case. In Riemannian geometry Weyl tensor is identically zero for any  homogeneous and isotropic cosmological model.

The nonzero rate of the dust particle production will contribute to both Weyl current  $G^\mu$ and energy-momentum tensor $T^{\mu\nu}$. The corresponding calculations are very cumbersome. We postpone their derivation and investigation to the near future.

\section{Conclusions and discussions}

We considered the homogeneous and isotropic cosmological models in the Weyl conformal geometry. This problem is by no means trivial since in the Riemannain conformal gravity all the homogeneous and isotropic metrics are the vacuum ones, i.\,e., the corresponding universes are empty. First of all, we noticed that the action integral for the matter field is not obliged to be conformal invariant, but its variation must have such a symmetry. From this we derived the corresponding self-consistency condition. It appears, that the matter content allows to have the dust whose energy-momentum tensor is nonzero. The requirement for this is the existence of direct dust particles interaction with the Weyl vector --- the part of Weyl geometry. We discovered that the number of dust particles is not conserved and found the general law for the rate of their production. We also found all vacuum cosmological solutions to the field equations with constant curvature. In this case the particles cannot be created at all. It is interesting to search for the vacuum solutions with evolving curvature, when particles can be, in principle, produced, but they are not still present We call such a situation ``{\it the pregnant vacuum}''.

\end{document}